\documentclass[a4paper,12pt]{article}

\def\displayandname#1{\rlap{$\displaystyle\csname #1\endcsname$}%
                      \qquad \texttt{\char92 #1}}

\usepackage{epsfig}

\begin{document}

\title{Classical Electrodynamics without Fields and the Aharonov-Bohm effect}

\author{Eugene V. Stefanovich \and \small \emph{2255 Showers Drive, Apt.
153, Mountain View, CA 94040, USA} \and \small
$eugene\_stefanovich@usa.net$}

\maketitle

\begin{abstract}
 The Darwin-Breit Hamiltonian is applied to the Aharonov-Bohm
 experiment. In agreement with the standard Maxwell-Lorentz theory, the force acting on electrons from
 infinite solenoids or ferromagnetic rods vanishes. However, the
interaction energies and phase factors of the electron wave packets
are non-zero. This allows us to explain the Aharonov-Bohm effect
 without involvement of electromagnetic potentials, fields, and
 topological properties of space.
\end{abstract}

\section{ Introduction} \label{sc:introduction}

In recent proposals to reformulate classical Maxwell-Lorentz
electrodynamics, electric and magnetic interactions propagate
instantaneously \cite{Chubykalo, Field-06} and the free
electromagnetic field is an approximation for a large ensemble of
discrete quantum particles - photons \cite{Field-photons, Torre3,
Carroll}. These ideas of "field-less" electrodynamics are attractive
for several reasons. First, the traditional continuous field
description of radiation is in conflict with corpuscular properties
of light that are evident in all kinds of single-photon experiments,
in the photo-electric effect, etc. Second, the traditional notions
of the momentum and energy contained in electromagnetic fields lead
to divergences, ``4/3 problem'' and other paradoxes
\cite{Rohrlich-60, Butler, Comay, Franklin}, which can be avoided in
the field-less description. Third, the usual assumption of the
retarded character of the Coulomb and magnetic interactions has not
been confirmed by experiment,\footnote{Of course, there exist
indirect interactions between particles transmitted by \emph{real}
photons emitted, absorbed, and scattered by accelerated charges
\cite{Chubykalo, Field-06, mybook}. It is well-established that
these indirect interactions propagate with the speed of light $c$,
and they are responsible for radar, radio, TV, etc. signals. We will
not discuss the electromagnetic radiation effects in this paper.}
and this assumption results in the paradox of energy
non-conservation \cite{Kislev}.\footnote{The explanation of this
paradox suggested in \cite{Kislev} is not satisfactory, because it
assumes interaction of overlapping light waves, which is known to be
negligibly small from QED.} On the other hand, the indirect support
for field-less electrodynamics is provided by numerous experiments
\cite{Giakos, superluminal1, superluminal2, superluminal3, Mugnai,
superluminal4, superluminal5, Kholmetskii}, which can be interpreted
as an evidence of instantaneous action-at-a-distance. It was also
demonstrated that such a superluminality does not violate the
principle of causality \cite{Stefanovich_Mink, mybook}, and that
action-at-a-distance
 potentials can be a reasonable alternative to the
general relativistic description of gravity
\cite{Stefanovich-gravity}.

First attempts at Hamiltonian formulations of electrodynamics
without fields were undertaken by Darwin \cite{Darwin} and Breit
\cite{Breit}. They found that electromagnetic effects in the
$(1/c)^2$ approximation can be represented by instantaneous
interparticle forces. The relativistic invariance of this approach
was established in \cite{Coleman}.  The Darwin-Breit Hamiltonian was
successfully applied to various electromagnetic problems, such as
the fine structure in atomic spectra \cite{BLP, Breitenberger},
superconductivity and properties of plasma \cite{Essen3, Essen1,
Essen2}.

The Aharonov-Bohm effect \cite{Aharonov, Chambers, Tonomura-86,
Tonomura-86a} is usually believed to be an indication of the
fundamental importance of electromagnetic potentials and fields in
nature. Although, several non-conventional explanations of this
effect were suggested in the literature \cite{Boyer-Darwin,
Spavieri-92, Hegerfeldt-08}, as far as I know, there were no
attempts to interpret this effect in terms of direct interactions
between particles. In this paper we would like to fill this gap and
to suggest a simple description of the Aharonov-Bohm effect within
the Darwin-Breit action-at-a-distance theory. This explanation does
not involve the notions of electromagnetic potentials, fields, and
non-trivial space topologies.

In section \ref{sc:hamiltonian} we briefly discuss the relativistic
Hamiltonian quantum mechanics and formulation of the Darwin-Breit
theory as a classical limit of the dressed particle version of
quantum electrodynamics (QED) \cite{Stefanovich_qft, mybook}.  In
section \ref{sc:two-particle} we apply the Darwin-Breit Hamiltonian
to interactions of a moving point charge with solenoids and
ferromagnets. The new approach to the Aharonov-Bohm effect is
discussed in section \ref{sc:ahar}.

\section {Relativistic Hamiltonian dynamics} \label{sc:hamiltonian}

 One class
of problems characteristic to Maxwell-Lorentz electrodynamics is
related to the apparent non-conservation of total observables
(energy, momentum, angular momentum, etc.) in systems of interacting
charges. Indeed, in the theory based on Maxwell's equations there is
no guarantee that total observables are conserved, that Newton's
third law of action and reaction is valid, and that total energy and
momentum form a 4-vector quantity. Suggested solutions of these
paradoxes \cite{Rohrlich-60, Keller, Page, Shockley, Furry,
Aharonov-88, Comay-hidden, Comay-hidden2, Jefimenko-99,
Kholmetskii-hidden, Hnizdo} involved such \emph{ad hoc}
constructions as ``hidden momentum'', the energy and momentum of
electromagnetic fields, "Poincar\'e stresses", etc.

In theoretical physics it is well established that conservation laws
are consequences of the invariance of observations with respect to
inertial transformations of reference frames. These transformations
are elements of the Poincar\'e group. So, the conservation of total
observables and their correct transformation properties can be
guaranteed in an approach based on a (Hamiltonian) theory of
representations of the Poincar\'e group. In quantum mechanics, such
an approach is realized within Wigner-Dirac theory
\cite{Wigner_unit, Dirac, book} in which dynamics of any isolated
physical system is described as a representation of the Poincar\'e
Lie algebra by Hermitian operators in the Hilbert space of states.
Representatives of ten basis elements (generators) of the Poincar\'e
Lie algebra are identified with observables of the total linear
momentum $\mathbf{P}$, total angular momentum $\mathbf{J}$, total
energy (the Hamiltonian) $H$, and "boost" operator\footnote{This
operator does not have interpretation as a common mechanical
observable, however it is closely related to observables of the
center-of-mass position and spin \cite{mybook}.} $\mathbf{K}$.  In
the instant form of Dirac's dynamics \cite{Dirac}, these generators
have the form

\begin{eqnarray}
\mathbf{P} &=& \mathbf{P}_0  \label{eq:P}\\
\mathbf{J} &=& \mathbf{J}_0  \label{eq:J} \\
\mathbf{K} &=& \mathbf{K}_0 + \mathbf{Z} \label{eq:k-z}\\
H &=& H_0 + V \label{eq:H}
\end{eqnarray}

\noindent where the non-interacting parts $\mathbf{P}_0,
\mathbf{J}_0, \mathbf{K}_0, H_0$ are simply sums of one-particle
generators, and interactions are contained in the \emph{potential
energy} $V$ and \emph{potential boost} $\mathbf{Z}$ operators.

 The commutator of any observable $F$ with the Hamiltonian $H$ determines
 the time evolution of this observable in the Heisenberg picture of
 quantum mechanics

\begin{eqnarray*}
\frac{dF(t)}{dt}  &=&  \frac{i}{\hbar} [H, F]
\end{eqnarray*}

\noindent Then the conservation of observables $H, \mathbf{P}$ and
$\mathbf{J}$ follows automatically from their vanishing commutators
with $H$. These conservation laws hold true independent on
interactions that may be present in the multiparticle system. This
simple fact is not at all obvious in the Maxwell-Lorentz theory,
which, for example, has serious difficulties in explaining the
conservation of the total angular momentum of a moving capacitor in
the Trouton-Noble experiment \cite{Butler, Furry, Page, Spavieri,
Teukolsky, Jefimenko-99a, Jackson-torque}.

 There are three essential steps \cite{mybook} that need
to be made in order to arrive at the Darwin-Breit Hamiltonian from
QED, which is rightly considered the most accurate physical theory
in existence. First, the "dressed particle" approach \cite{GS,
Shirokov4, Stefanovich_qft, mybook} should be applied, which allows
one to formulate the quantum field theory in terms of physical
(rather than "bare") particles
 and avoid ultraviolet divergences. This leads
to the perturbation expansion of the potential energy operator $V$
whose terms are direct particle interactions. In the second
perturbation order the interaction is a sum of two-particle terms,
so  it is possible to consider only the two-particle sector of the
Fock space and express the interaction energy as a function of
particle charges $q_i$, positions $\mathbf{r}_i$, momenta
$\mathbf{p}_i$, spins $\mathbf{s}_i$, and energies $h_i =
\sqrt{m_i^2c^4 + p_i^2c^2}$. Second, in the classical limit ($\hbar
\to 0$) commutators of operators are replaced by \emph{Poisson
brackets} $\{\ldots,\ldots \}$. Finally, for low-velocity processes
of classical electrodynamics one can represent all quantities as
series in powers of $1/c$ and leave only terms of order not higher
than $(1/c)^{2}$. In these approximations
 the set
of generators (\ref{eq:P}) - (\ref{eq:H}) takes the form

\begin{eqnarray}
\mathbf{P}_0 &=& \mathbf{p}_1 + \mathbf{p}_2 \nonumber \\
\mathbf{J}_0 &=& [\mathbf{r}_1 \times \mathbf{p}_1]
+ \mathbf{s}_1 + [\mathbf{r}_2 \times \mathbf{p}_2] + \mathbf{s}_2 \nonumber \\
H_0 &=& h_1 + h_2 \nonumber
\\
&\approx& m_1c^2 + m_2c^2 + \frac{p_1^2}{2m_1} +
\frac{p_2^2}{2m_2}-
\frac{p_1^4}{8m_1^3c^2}   - \frac{p_2^4}{8m_2^3c^2}\label{eq:mc2} \\
 V &\approx&  V_{Coulomb} + V_{Darwin} + V_{spin-orb} +
V_{spin-spin}  \label{eq:full-H} \nonumber \\
V_{Coulomb} &=& \frac{q_1q_2}{4 \pi r} \label{eq:coulomb} \\
V_{Darwin} &=&  - \frac{q_1q_2}{8 \pi m_1m_2c^2 r}
\left((\mathbf{p_1} \cdot \mathbf{p}_2) + \frac{(\mathbf{p_1} \cdot
\mathbf{r})(\mathbf{p_2}
\cdot \mathbf{r})}{r^2} \right) \label{eq:darwin} \\
 V_{spin-orbit} &=& \frac{q_1 ([\vec{\mu}_2
\times \mathbf{r}] \cdot (\mathbf{p}_2 - 2 \mathbf{p}_1))}{8 \pi m_2
c r^3} -
 \frac{ q_2  ([\vec{\mu}_1 \times \mathbf{r} ]
\cdot (\mathbf{p}_1 - 2 \mathbf{p}_2))}{8 \pi m_1 c r^3}
\label{eq:H-spin-orb}\\
 V_{spin-spin}
&=&  \frac{(\vec{\mu}_2 \cdot \vec{\mu}_1)} {4 \pi
 r^3} - \frac{3(\vec{\mu}_2 \cdot \mathbf{r}) (\vec{\mu}_1 \cdot
\mathbf{r})}{4 \pi  r^5} \label{eq:H-spin-spin} \\
\mathbf{K}_0 &=& -\frac{h_1\mathbf{r}_1}{c^2} - \frac{[\mathbf{p}_1
\times \mathbf{s}_1]}{m_1c^2 + h_1} -\frac{h_2\mathbf{r}_2}{c^2} -
\frac{[\mathbf{p}_2 \times
\mathbf{s}_2]}{m_2c^2 + h_2} \nonumber \\
&\approx& -m_1 \mathbf{r}_1 - m_2 \mathbf{r}_2 -\frac{p_1^2
\mathbf{r}_1}{2m_1c^2}  - \frac{p_2^2 \mathbf{r}_2}{2m_2 c^2}  +
\frac{1}{2 c^2 } \left(\frac{[\mathbf{s}_1 \times \mathbf{p}_1
]}{m_1} +
\frac{[\mathbf{s}_2 \times \mathbf{p}_2 ]}{m_2} \right) \nonumber\\
\mathbf{Z} &\approx&    -  \frac{q_1q_2(\mathbf{r}_1 +
\mathbf{r}_2)}{8 \pi c^2r} \label{eq:boost-spin-orb} \nonumber
\end{eqnarray}

\noindent The Hamiltonian $H = H_0 +V$ is the Darwin-Breit
Hamiltonian for two charged spinning particles.\footnote{We use the
Heaviside-Lorentz system of units  and denote $\mathbf{r} \equiv
\mathbf{r}_1 - \mathbf{r}_2$ throughout this paper. This form of the
Darwin-Breit Hamiltonian can be found in eqs. (9.50) - (9.55) in
\cite{mybook}, where contact terms proportional to
$\delta(\mathbf{r})$ are not relevant for classical mechanics and
can be omitted. Also here we express the potential energy of
interaction in terms of particles' magnetic moments $ \vec{\mu}_i =
e\mathbf{s}_i/(m_i c)$ which correspond to the gyromagnetic ratio $g
\approx 2$ characteristic for electrons.} It
 contains the familiar Coulomb
energy (\ref{eq:coulomb}) and the \emph{Darwin potential} energy
(\ref{eq:darwin}) \cite{Darwin}, which is responsible for magnetic
interactions between charged particles.  Further relativistic
corrections are given by the spin-orbit (\ref{eq:H-spin-orb}) and
spin-spin (\ref{eq:H-spin-spin}) interactions. A straightforward
computation shows that Poisson brackets of the above generators
satisfy the Poincar\'e Lie algebra relationships within $(1/c)^2$
approximation \cite{Coleman, Close-Osborn, Krajcik-Foldy}.

\section{Interaction between charges and magnetic dipoles}
\label{sc:two-particle}

In the Aharonov-Bohm effect \cite{Aharonov} a charged particle
(e.g., an electron labeled by the index 1) interacts with an
infinite solenoid or ferromagnetic rod (which can be represented as
a collection of point magnetic moments 2). This experiment is not
sensitive to the electron's spin orientation, and the charge of the
solenoid/rod is zero. Then we can assume $\vec{\mu}_1 = 0$ and
$q_2=0$, omit interaction terms (\ref{eq:coulomb}),
(\ref{eq:darwin}), (\ref{eq:H-spin-spin}), and write the simplified
Hamiltonian for the system "point magnetic dipole +
charge"\footnote{Here we drop the first two terms in (\ref{eq:mc2}),
which are the rest energies of the two particles and do not have any
effect on dynamics.}

\begin{eqnarray}
 H
&=& \frac{p_1^2}{2m_1} + \frac{p_2^2}{2m_2} -
\frac{p_1^4}{8m_1^3c^2}   - \frac{p_2^4}{8m_2^3c^2} + \frac{q_1
[\vec{\mu}_2 \times \mathbf{r}] \cdot \mathbf{p}_2}{8\pi m_2 c r^3}
- \frac{q_1 [\vec{\mu}_2 \times \mathbf{r}] \cdot \mathbf{p}_1}{4
\pi m_1c r^3} \label{eq:h-charge-spin}
\end{eqnarray}

The time derivative of the first particle's momentum can be obtained
from the Hamilton's equation of motion

\begin{eqnarray*}
\frac{d \mathbf{p}_1}{dt} &=& \{H, \mathbf{p}_1\} = - \frac{\partial
H}{\partial
\mathbf{r}_1} \\
&=&
 \frac{q_1[ \mathbf{p}_1 \times \vec{\mu}_2] }{4 \pi m_1 cr^3}
 -
 \frac{3q_1 ([ \mathbf{p}_1 \times \vec{\mu}_2] \cdot \mathbf{r})\mathbf{r} }{4\pi m_1 cr^5}
 - \frac{q_1
[\mathbf{p}_2 \times \vec{\mu}_2]  }{8\pi m_2 c r^3} +  \frac{3q_1
([\mathbf{p}_2 \times \vec{\mu}_2] \cdot \mathbf{r}) \mathbf{r}
}{8\pi m_2 c r^5} \label{eq:dp1dt}
\end{eqnarray*}

\noindent The time derivative  of the second particle's momentum
follows from the law of conservation of the total momentum
($[\mathbf{P}, H] = 0$)

\begin{eqnarray}
 \frac{d \mathbf{p}_2}{dt} &=& \{\mathbf{p}_2,H\} = \{\mathbf{P}
-\mathbf{p}_1,H \} = -\{\mathbf{p}_1,H\} = -\frac{d
\mathbf{p}_1}{dt} \label{eq:dp2/dt}
\end{eqnarray}

\noindent This is the third Newton's law of action and reaction,
which holds exactly in the instant form of dynamics, and there is no
need to invoke such dubious notions as ``hidden momentum''  and/or
momentum of electromagnetic fields in order to enforce this law.

It is difficult to measure momenta of particles and their time
derivatives in experiment. It is much easier to measure velocities
and accelerations, e.g., by the time-of-flight technique
\cite{Caprez}. The velocity of the charged particle 1 is obtained
from the Hamilton's equation

\begin{eqnarray}
\mathbf{v}_1 \equiv \frac{d \mathbf{r}_1}{dt} = \{\mathbf{r}_1, H \}
= \frac{\partial H}{\partial \mathbf{p}_1} =
\frac{\mathbf{p}_1}{m_1}- \frac{p_1^2\mathbf{p}_1}{2m_1^3 c}  -
\frac{q_1 [\vec{\mu}_2 \times \mathbf{r}] }{4 \pi m_1c r^3}
\label{eq:dr1dt}
\end{eqnarray}

\noindent  This relationship  is interaction-dependent because the
interaction energy in (\ref{eq:h-charge-spin}) is
momentum-dependent. From (\ref{eq:dp1dt}), (\ref{eq:dr1dt}), and
vector identity $[\mathbf{a} \times [ \mathbf{b} \times \mathbf{c}]]
= \mathbf{b}(\mathbf{a} \cdot \mathbf{c}) - \mathbf{c}(\mathbf{a}
\cdot \mathbf{b})$ we obtain the acceleration of the charged
particle interacting with the magnetic moment at rest ($\mathbf{p}_2
= 0$)

\begin{eqnarray}
\mathbf{a}_1 &\equiv& \frac{d^2 \mathbf{r}_1}{dt^2} =
\{\mathbf{v}_1, H
\} \nonumber \\
 &\approx& \frac{\dot{\mathbf{p}}_1}{m_1}-
 \frac{q_1[  \vec{\mu}_2 \times \dot{\mathbf{r}}] }{4 \pi m_1^2c r^3} +
 \frac{3q_1[  \vec{\mu}_2 \times \mathbf{r}] (\mathbf{r} \cdot \dot{\mathbf{r}})}{4 \pi m_1^2c r^5}
 \nonumber \\
 &=& \frac{q_1[ \mathbf{p}_1 \times \vec{\mu}_2] }{2 \pi m_1^2 cr^3}
 -
 \frac{3q_1 ([ \mathbf{p}_1 \times \vec{\mu}_2] \cdot \mathbf{r})\mathbf{r} }{4 \pi m_1^2 cr^5}
 +
 \frac{3q_1[  \vec{\mu}_2 \times \mathbf{r}] (\mathbf{r} \cdot \mathbf{p}_1)}{4 \pi m_1^2 cr^5}
 \nonumber \\
  &=& \frac{q_1 [\mathbf{p}_1 \times \vec{\mu}_2]}{2
\pi m_1^2 cr^3} -
 \frac{3q_1 [ \mathbf{p}_1 \times [\mathbf{r} \times [\vec{\mu}_2 \times \mathbf{r}]]]
 }{4 \pi m_1^2c  r^5} \nonumber \\
 &=& -\frac{q_1 [\mathbf{p}_1 \times \vec{\mu}_2]}{4 \pi
m_1^2 cr^3} +
 \frac{3q_1 [ \mathbf{p}_1 \times \mathbf{r}](\vec{\mu}_2 \cdot
 \mathbf{r})
 }{4 \pi m_1^2c  r^5} \nonumber \\
  &\approx& \frac{q_1}{m_1 c}[\mathbf{v}_1 \times \mathbf{B}]
  \label{eq:lorentz-force}
\end{eqnarray}

\noindent   This agrees with the standard Lorentz force formula if
another standard expression (see eq. (5.56) in \cite{Jackson})

\begin{eqnarray}
 \mathbf{B} = - \frac{\vec{\mu}_2}{4 \pi   r^3} + \frac{3(\vec{\mu}_2 \cdot \mathbf{r}) \mathbf{r}}{4 \pi
r^5} \label{eq:field}
\end{eqnarray}

\noindent  is used for the "magnetic field" of the magnetic
moment.\footnote{We write "magnetic field" in quotes, because in the
Darwin-Breit approach there are no fields (electric or magnetic)
having independent existence at each space point. There are only
direct inter-particle forces, and in eq. (\ref{eq:field})
$\mathbf{r}_1$ and $\mathbf{r}_2$ are coordinates of two particles,
rather than general points in space. }

In Appendix A we show that the same "magnetic field" is created by a
small circular loop with current.  The "magnetic field" of
infinitely long thin solenoid or infinitely long ferromagnetic rod
can be obtained by integrating (\ref{eq:field}) along the length of
the solenoid/rod. It is easy to show that this integral vanishes.
This agrees with the prediction of Maxwell's theory that a charge is
moving without acceleration in the vicinity of an infinite
magnetized solenoid/rod. In particular, this result is consistent
with the lack of "time lag" in experiments \cite{Caprez}.

\begin{figure}
 \epsfig {file=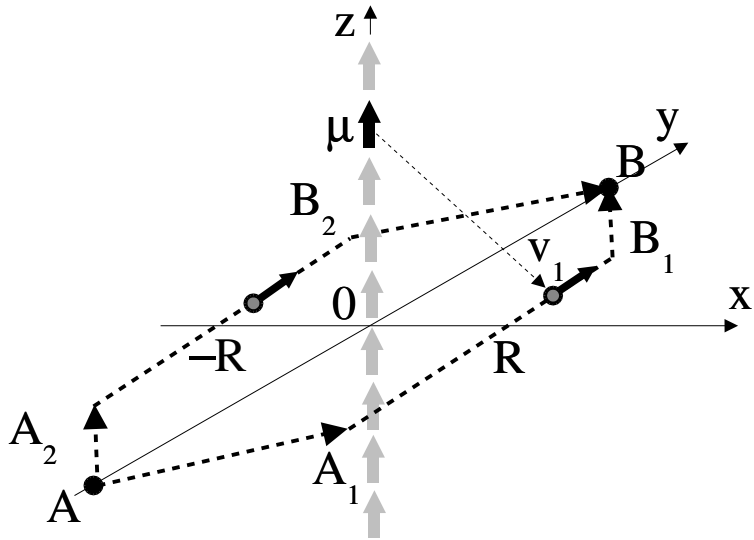} \caption{The Aharonov-Bohm experiment. } \label{fig:4}
\end{figure}

\section {The Aharonov-Bohm effect} \label{sc:ahar}

Let us consider the following idealized version of the Aharonov-Bohm
experiment (see fig. \ref{fig:4}): An infinite solenoid or
 ferromagnetic rod with negligible cross-section and linear magnetization $\mu$
 is erected vertically in the origin
 (grey arrows). The electron wave packet is split into two parts (e.g., by using a double-slit)
 at point $A$. These subpackets travel on both sides of the solenoid/rod
 with constant velocity $\mathbf{v}_1$, and the distance of the
 closest approach is $R$.
 The subpackets  rejoin at point $B$, where the interference is measured.
 (The two trajectories $AA_1B_1B$ and
 $AA_2B_2B$ are denoted by broken lines.)
 The distance $AB$ is sufficiently large, so that electron's path
can be assumed parallel to the $y$-axis everywhere

\begin{eqnarray}
\mathbf{r}_1(t) = (\pm R, v_1t, 0) \label{eq:rt}
\end{eqnarray}

 Experimentally it
was found  that the interference of the two wave packets at point
$B$ depends on the magnetization of the solenoid/rod
\cite{Chambers}. In the preceding section we demonstrated that
electron's acceleration is zero. Therefore the Aharonov-Bohm effect
cannot be explained as a result of classical forces
\cite{Boyer-Darwin, Boyer-Ahar, Boyer-2007-08, Boyer-2007}. To
resolve this paradox it is sufficient to mention that the
representation of the wave packet as a point moving through space
along the trajectory (\ref{eq:rt}) is an oversimplification. A more
complete description of the electron's wave function should also
include the overall phase factor

\begin{eqnarray*}
\psi_1 (\mathbf{r},t) \approx e^{\frac{i}{\hbar}S(t)} \delta
(\mathbf{r} - \mathbf{r}_1(t)) \label{eq:rps}
\end{eqnarray*}

\noindent The \emph{action} integral $S(t)$ for the one-particle
wave packet that traveled between time points $t_0$ and $t$ is

\begin{eqnarray}
S(t) &\equiv& \int \limits_{t_0}^{t} \left(\frac{ m_1 v_1^2(t')}{2}
- V_1(t') \right) dt' \label{eq:phase-factor}
\end{eqnarray}

\noindent where $V_1(t)$ is the contribution to the particle's
energy due to the external potential. Then the interference of the
"left" and "right" wave packets at point $B$ will depend on the
relative value of phase factors accumulated by them along the path
$AB$

\begin{eqnarray*}
\phi &=& \frac{1}{\hbar}(S_{right} - S_{left}) \label{eq:right}
\end{eqnarray*}

Let us now calculate the relative phase shift in the geometry of
fig. \ref{fig:4}. The kinetic energy term in (\ref{eq:phase-factor})
does not contribute, because velocity remains constant for both
paths. However, the potential energy of the charge\footnote{Here we
integrate the last term in eq. (\ref{eq:h-charge-spin}) along the
length of the solenoid and notice that the mixed product
$([\vec{\mu} \times \mathbf{v}_1 ] \cdot \mathbf{r}_1)$ is
independent on $z$.}

\begin{eqnarray*}
V =  \int \limits_{-\infty}^{\infty} dz \frac{q_1 ([\vec{\mu} \times
\mathbf{v}_1 ] \cdot \mathbf{r}_1)}{4 \pi  c (x^2+y^2+z^2)^{3/2}} =
\frac{q_1 ([\vec{\mu} \times \mathbf{v}_1 ] \cdot \mathbf{r}_1)}{2
\pi c (x^2+y^2)} \label{eq:v-solen}
\end{eqnarray*}

\noindent is different for the two paths. For all points on the
"right" path the numerator of this expression is $ - q_1\mu v_1 R$,
and for the "left" path the numerator is $q_1\mu v_1 R$. Then the
total phase shift is

\begin{eqnarray*}
\phi &=& \frac{1}{\hbar} \int \limits_{-\infty}^{\infty}  \frac{
q_1\mu Rv_1}{ \pi  c (R^2 + v_1^2t^2)} dt  =  \frac{e\mu }{\hbar c}
\end{eqnarray*}

\noindent This phase difference does not depend on the electron's
velocity and on the value of $R$. However, it is proportional to the
rod's magnetization
 $\mu$.   So, all essential properties of the
Aharonov-Bohm effect are fully described within the Darwin-Breit
direct interaction theory.\footnote{These results were derived for
 thin ferromagnetic rods and solenoids, however the same arguments
apply to cylindrical rods and solenoids of any cross-section. A
cylindrical rod can be represented as a bunch of thin rods. A
cylindrical solenoid also can be represented as a bunch of thin
solenoids. In this representation, the currents cancel out in the
interior, where neighboring components touch each other, and only
currents on the outside surface have effect on the charge $q_1$. The
same phase shift formula can be obtained for toroidal solenoids,
which were used in Tonomura's experiments \cite{Tonomura-86,
Tonomura-86a}. } In our description the Aharonov-Bohm effect is a
quantum phenomenon, however, in contrast to traditional views, this
effect does not prove the existence of scalar and/or vector
electromagnetic potentials, and it is not essential whether the
solenoid/rod is infinite (so that it induces a multiple-connected
topology of space) or not. The latter point is supported by
experiments with finite-length magnetized nanowires, which exhibit
the phase shift similar to that characteristic for infinite
solenoids/rods \cite{Matteucci}.

I am thankful to Dr. Peter Enders for helpful comments and
discussions.

\appendix

\section {Appendix. Current loop and charge}

\begin{figure}
 \epsfig {file=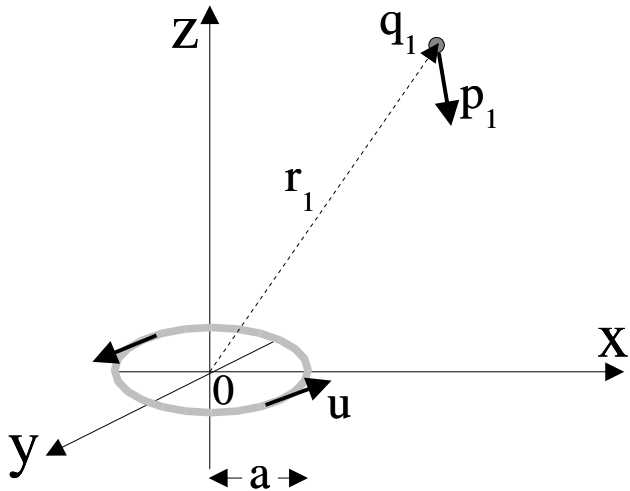} \caption{Interaction between current loop and charge}
\label{fig:1}
\end{figure}

Let us calculate the interaction energy between a
 neutral circular current loop of small radius $a$  and a
point charge in the geometry shown in fig. \ref{fig:1}. There are
three types of charges in this problem: First, there is the charge
$q_1$ located at a general point in space $\mathbf{r}_1 =
(r_{1x},r_{1y},r_{1z})$ and having arbitrary momentum $\mathbf{p}_1
= (p_{1x},p_{1y},p_{1z})$. Also there are positive charges of
immobile ions in the metal uniformly distributed along the loop with
linear density $\rho_3$ and negative charges of conduction electrons
having linear density $ \rho_2 = -\rho_3 $.

Let us introduce a few simplification in this problem. First, we are
not interested in the interaction between charge densities $\rho_2$
and $\rho_3$. Second, the Coulomb interactions of charges $1
\leftrightarrow 2$ and $1 \leftrightarrow 3$ cancel each other.
Third, it can be shown that if the current loop moves with total
velocity $\mathbf{v}$, then $\mathbf{v}$-dependent interactions
(\ref{eq:darwin}) of the  charge 1 with electrons and ions in the
loop also cancel each other. Therefore, we can assume that the
current loop is stationary in the origin, and that only electrons in
the loop are moving with velocity $\mathbf{v}_2 \approx
\mathbf{p}_2/m_2$ whose tangential component is $u$, as shown in
fig. \ref{fig:1}. Finally, the spins of ions and electrons in the
wire are oriented randomly, therefore Hamiltonian terms
(\ref{eq:H-spin-orb}) - (\ref{eq:H-spin-spin}), being linear with
respect to particle spins, vanish after averaging. Then the
potential energy of interaction between the charge 1 and the loop
element $dl$ is given by the Darwin's formula

\begin{eqnarray}
 V_{dl-q1}
&\approx& -\frac{q_1\rho_2 dl}{8 \pi m_1 c^2}
\left(\frac{(\mathbf{p}_1 \cdot \mathbf{v}_2)}{r} +
\frac{(\mathbf{p}_1 \cdot \mathbf{r})( \mathbf{v}_2 \cdot \mathbf{r}
) }{r^3} \right) \nonumber
\end{eqnarray}

\noindent In the coordinate system shown in fig. \ref{fig:1} the
line element in the loop is $dl = a d \theta$ and $\mathbf{v}_2 =
(-u \sin \theta, u \cos \theta, 0)$. In the limit $a \to 0$ we can
approximate

\begin{eqnarray*}
r^{-1}
&\approx&  \frac{1}{r_1} + \frac{a (r_{1x}\cos \theta + r_{1y}\sin \theta)}{r_1^3} \\
r^{-3} &\approx& \frac{1}{r_1^3} + \frac{3a (r_{1x}\cos \theta +
r_{1y}\sin \theta)}{r_1^5}
\end{eqnarray*}

\noindent The full interaction between the charge and the loop is
obtained by integrating $V_{dl-q1}$ on $\theta$ from 0 to $2 \pi$
and neglecting small terms proportional to $a^3$

\begin{eqnarray*}
&\ & V_{loop-q1} \nonumber \\
&\approx&  - \frac{aq_1 \rho_2}{8 \pi m_1 c^2} \int \limits _{0}^{2
\pi} d \theta \Bigl[(-up_{1x} \sin \theta+ u p_{1y}\cos \theta)
 \left(\frac{1}{r_1} + \frac{a (r_{1x}\cos \theta + r_{1y}\sin \theta)}{r_1^3} \right) \nonumber \\
 &+& (-ur_{1x} \sin \theta+ u r_{1y}\cos \theta)(
(\mathbf{p}_1 \cdot \mathbf{r}_1) - p_{1x}a \cos \theta- p_{1y} a
\sin \theta ) \times \\
&\ & \left(\frac{1}{r_1^3} + \frac{3a (r_{1x}\cos \theta +
r_{1y}\sin \theta )}{r_1^5} \right) \Bigr]
\nonumber \\
&\approx&  -\frac{a^2uq_1 \rho_2 [\mathbf{r}_1 \times
\mathbf{p}_1]_z}{4 m_1 c^2 r_1^3} \label{eq:vloop-q1}
\end{eqnarray*}

\noindent  Taking into account the usual definition of the loop's
magnetic moment $ \mu_2 = \pi a^2 \rho_2 u/c$ (see eq. (5.42) in
\cite{Jackson}) whose direction is orthogonal to the plane of the
loop, we find that for arbitrary position and orientation of the
loop

\begin{eqnarray*}
V_{loop-q1} &\approx&  -
 \frac{ q_1[\vec{\mu}_2 \times \mathbf{r}]
\cdot \mathbf{p}_1}{4 \pi m_1 c r^3} \label{eq:vloopq1}
\end{eqnarray*}

\noindent which agrees with the spin-charge interaction in
(\ref{eq:H-spin-orb}) when $\mathbf{p}_2=0$. Therefore, the
acceleration of the charge $q_1$ moving in the field of the current
loop is also given by eq. (\ref{eq:lorentz-force}).


\begin{thebibliography}{10}

\bibitem{Chubykalo}
A.~E. Chubykalo,  R.~Smirnov-Rueda.
\newblock Action at a distance as a full-value solution of {Maxwell} equations:
  basis and application of separated potential's method.
\newblock {\em Phys. Rev. E}, \textbf{53}:5373, 1996.

\bibitem{Field-06}
J.~H. Field.
\newblock Classical electromagnetism as a consequence of {Coulomb's} law,
  special relativity and {Hamilton's} principle and its relationship to quantum
  electrodynamics.
\newblock {\em Phys. Scr.}, \textbf{74}:702, 2006.
\newblock http://www.arxiv.org/abs/physics/0501130v5.

\bibitem{Field-photons}
J.~H. Field.
\newblock On the relationship of quantum mechanics to classical
  electromagnetism and classical relativistic mechanics.
\newblock http://www.arxiv.org/abs/physics/0403076.

\bibitem{Torre3}
A.~C. de~la Torre.
\newblock Understanding light quanta: {Construction} of the free
  electromagnetic field.
\newblock http://www.arxiv.org/abs/quant-ph/0503023v2.

\bibitem{Carroll}
R.~Carroll.
\newblock Remarks on photons and the aether.
\newblock http://www.arxiv.org/abs/physics/0507027v1.

\bibitem{Rohrlich-60}
F.~Rohrlich.
\newblock Self-energy and stability of the classical electron.
\newblock {\em Am. J. Phys.}, \textbf{28}:639, 1960.

\bibitem{Butler}
J.~W. Butler.
\newblock A proposed electromagnetic momentum-energy 4-vector for charged
  bodies.
\newblock {\em Am. J. Phys.}, \textbf{37}:1258, 1969.

\bibitem{Comay}
E.~Comay.
\newblock Decomposition of electromagnetic fields into radiation and bound
  components.
\newblock {\em Am. J. Phys.}, \textbf{65}:862, 1997.

\bibitem{Franklin}
J.~Franklin.
\newblock The nature of electromagnetic energy, 2007.
\newblock http://www.arxiv.org/abs/0707.3421v2.

\bibitem{mybook}
E.~V. Stefanovich.
\newblock Relativistic quantum dynamics, 2005.
\newblock http://www.arxiv.org/abs/physics/0504062v7.

\bibitem{Kislev}
A.~Kislev,  L.~Vaidman.
\newblock Relativistic causality and conservation of energy in classical
  electromagnetic theory.
\newblock http://www.arxiv.org/abs/physics/0201042v1.

\bibitem{Giakos}
G.~C. Giakos,  T.~K. Ishii.
\newblock Rapid pulsed microwave propagation.
\newblock {\em IEEE Microwave and Guided Wave Letters}, \textbf{1}:374, 1991.

\bibitem{superluminal1}
A.~Enders,  G.~Nimtz.
\newblock Evanescent-mode propagation and quantum tunneling.
\newblock {\em Phys. Rev. E}, \textbf{48}:632, 1993.

\bibitem{superluminal2}
A.~M. Steinberg, P.~G. Kwiat, R.~Y. Chiao.
\newblock Measurement of the single-photon tunneling time.
\newblock {\em Phys. Rev. Lett.}, \textbf{71}:708, 1993.

\bibitem{superluminal3}
A.~Ranfagni,  D.~Mugnai.
\newblock Anomalous pulse delay in microwave propagation: {A} case of
  superluminal behavior.
\newblock {\em Phys. Rev. E}, \textbf{54}:5692, 1996.

\bibitem{Mugnai}
D.~Mugnai, A.~Ranfagni, R.~Ruggeri.
\newblock Observation of superluminal behaviors in wave propagation.
\newblock {\em Phys. Rev. Lett.}, \textbf{21}:4830, 2000.

\bibitem{superluminal4}
K.~Wynne,  D.~A. Jaroszynski.
\newblock Superluminal terahertz pulses.
\newblock {\em Optics Letters}, \textbf{24}:25, 1999.

\bibitem{superluminal5}
W.~D. Walker.
\newblock Experimental evidence of near-field superluminally propagating
  electromagnetic fields.
\newblock http://www.arxiv.org/abs/physics/0009023v1.

\bibitem{Kholmetskii}
A.~L. Kholmetskii, O.~V. Missevitch, R.~Smirnov-Rueda, R.~I. Tzonchev, A.~E.
  Chubykalo, I.~Moreno.
\newblock Experimental evidence on non-applicability of the standard
  retardation condition to bound magnetic fields and on new generalized
  {Biot-Savart} law.
\newblock {\em J. Appl. Phys.}, \textbf{101}:023532, 2007.
\newblock http://www.arxiv.org/abs/physics/0601084v1.

\bibitem{Stefanovich_Mink}
E.~V. Stefanovich.
\newblock Is {Minkowski} space-time compatible with quantum mechanics?
\newblock {\em Found. Phys.}, \textbf{32}:673, 2002.

\bibitem{Stefanovich-gravity}
E.~V. Stefanovich.
\newblock A {Hamiltonian} approach to quantum gravity, 2006.
\newblock http://www.arxiv.org/abs/physics/0612019v9.

\bibitem{Darwin}
C.~G. Darwin.
\newblock The dynamical motions of charged particles.
\newblock {\em Phil. Mag.}, \textbf{39}:537, 1920.

\bibitem{Breit}
G.~Breit.
\newblock The effect of retardation on the interaction of two electrons.
\newblock {\em Phys. Rev.}, \textbf{34}:553, 1929.

\bibitem{Coleman}
S.~Coleman,  J.~H.~Van Vleck.
\newblock Origin of {"hidden momentum forces"} on magnets.
\newblock {\em Phys. Rev.}, \textbf{171}:1370, 1968.

\bibitem{BLP}
V.~B. Berestetski{\u{\i}}, E.~M. Livshitz, L.~P. Pitaevski{\u{\i}}.
\newblock {\em Quantum electrodynamics}.
\newblock Fizmatlit, Moscow, 2001.
\newblock (in Russian).

\bibitem{Breitenberger}
E.~Breitenberger.
\newblock Magnetic interactions between charged particles.
\newblock {\em Am. J. Phys.}, \textbf{36}:505, 1968.

\bibitem{Essen3}
H.~Ess\'en.
\newblock A study of lattice and magnetic interactions of conduction electrons.
\newblock {\em Phys. Scr.}, \textbf{52}:388, 1995.

\bibitem{Essen1}
H.~Ess\'en.
\newblock Darwin magnetic interaction energy and its macroscopic consequences.
\newblock {\em Phys. Rev. E}, \textbf{53}:5228, 1996.

\bibitem{Essen2}
H.~Ess\'en.
\newblock Magnetism of matter and phase space energy of charged particle
  systems.
\newblock {\em J. Phys. A: Math. Gen.}, \textbf{32}:2297, 1999.

\bibitem{Aharonov}
Y.~Aharonov,  D.~Bohm.
\newblock Significance of electromagnetic potentials in quantum mechanics.
\newblock {\em Phys. Rev.}, \textbf{115}:485, 1959.

\bibitem{Chambers}
R.~G. Chambers.
\newblock Shift of an electron interference pattern by enclosed magnetic flux.
\newblock {\em Phys. Rev. Lett.}, \textbf{5}:3, 1960.

\bibitem{Tonomura-86}
A.~Tonomura, N.~Osakabe, T.~Matsuda, T.~Kawasaki, J.~Endo, S.~Yano, H.~Yamada.
\newblock Evidence for {Aharonov-Bohm} effect with magnetic field completely
  shielded from electron wave.
\newblock {\em Phys. Rev. Lett.}, \textbf{56}:792, 1986.

\bibitem{Tonomura-86a}
N.~Osakabe, T.~Matsuda, T.~Kawasaki, J.~Endo, A.~Tonomura, S.~Yano, H.~Yamada.
\newblock Experimental confirmation of {Aharonov-Bohm} effect using a toroidal
  magnetic field confined by a superconductor.
\newblock {\em Phys. Rev. A: Math. Gen.}, \textbf{34}:815, 1986.

\bibitem{Boyer-Darwin}
T.~H. Boyer.
\newblock {Darwin-Lagrangian} analysis for the interaction of a point charge
  and a magnet: {Considerations} related to the controversy regarding the
  {Aharonov-Bohm} and {Aharonov-Casher} phase shifts.
\newblock {\em J. Phys. A:Math. Gen.}, \textbf{39}:3455, 2006.
\newblock http://www.arxiv.org/abs/physics/0506181v1.

\bibitem{Spavieri-92}
G.~Spavieri,  G.~Cavalleri.
\newblock Interpretation of the {Aharonov-Bohm} and the {Aharonov-Casher}
  effects in terms of classical electromagnetic fields.
\newblock {\em Europhys. Lett.}, \textbf{18}:301, 1992.

\bibitem{Hegerfeldt-08}
G.~C. Hegerfeldt,  J.~T. Neumann.
\newblock The {Aharonov-Bohm} effect: the role of tunneling and associated
  forces, 2008.
\newblock http://www.arxiv.org/abs/arXiv:0801.0799v2.

\bibitem{Stefanovich_qft}
E.~V. Stefanovich.
\newblock Quantum field theory without infinities.
\newblock {\em Ann. Phys. (NY)}, \textbf{292}:139, 2001.

\bibitem{Keller}
J.~M. Keller.
\newblock Newton's third law and electrodynamics.
\newblock {\em Am. J. Phys.}, \textbf{10}:302, 1942.

\bibitem{Page}
L.~Page,  N.~I.~Adams Jr.
\newblock Action and reaction between moving charges.
\newblock {\em Am. J. Phys.}, \textbf{13}:141, 1945.

\bibitem{Shockley}
W.~Shockley,  R.~P. James.
\newblock "{Try} simplest cases" discovery of "hidden momentum" forces on
  "magnetic currents".
\newblock {\em Phys. Rev. Lett.}, \textbf{18}:876, 1967.

\bibitem{Furry}
W.~H. Furry.
\newblock Examples of momentum distributions in the electromagnetic field and
  in matter.
\newblock {\em Am. J. Phys.}, \textbf{37}:621, 1969.

\bibitem{Aharonov-88}
Y.~Aharonov, P.~Pearle, L.~Vaidman.
\newblock Comment on {"Proposed Aharonov-Casher} effect: {Another} example of
  an {Aharonov-Bohm} effect arising from a classical lag".
\newblock {\em Phys. Rev. A}, \textbf{37}:4052, 1988.

\bibitem{Comay-hidden}
E.~Comay.
\newblock Exposing "hidden momentum".
\newblock {\em Am. J. Phys.}, \textbf{64}:1028, 1996.

\bibitem{Comay-hidden2}
E.~Comay.
\newblock Lorentz transformation of a system carrying "hidden momentum".
\newblock {\em Am. J. Phys.}, \textbf{68}:1007, 2000.

\bibitem{Jefimenko-99}
O.~D. Jefimenko.
\newblock A relativistic paradox seemingly violating conservation of momentum
  law in electromagnetic systems.
\newblock {\em Eur. J. Phys.}, \textbf{20}:39, 1999.

\bibitem{Kholmetskii-hidden}
A.~L. Kholmetskii.
\newblock On momentum and energy of a non-radiating electromagnetic field,
  2005.
\newblock http://www.arxiv.org/abs/physics/0501148v2.

\bibitem{Hnizdo}
V.~Hnizdo.
\newblock On linear momentum in quasistatic electromagnetic systems, 2004.
\newblock http://www.arxiv.org/abs/physics/0407027v1.

\bibitem{Wigner_unit}
E.~P. Wigner.
\newblock On unitary representations of the inhomogeneous {Lorentz} group.
\newblock {\em Ann. Math.}, \textbf{40}:149, 1939.

\bibitem{Dirac}
P.~A.~M. Dirac.
\newblock Forms of relativistic dynamics.
\newblock {\em Rev. Mod. Phys.}, \textbf{21}:392, 1949.

\bibitem{book}
S.~Weinberg.
\newblock {\em The Quantum Theory of Fields, Vol. 1}.
\newblock University Press, Cambridge, 1995.

\bibitem{Spavieri}
G.~Spavieri,  G.~T. Gillies.
\newblock Fundamental tests of electrodynamic theories: {Conceptual}
  investigations of the {Trouton-Noble} and hidden momentum effects.
\newblock {\em Nuovo Cim.}, \textbf{118B}:205, 2003.

\bibitem{Teukolsky}
S.~A. Teukolsky.
\newblock The explanation of the {Trouton-Noble} experiment revisited.
\newblock {\em Am. J. Phys.}, \textbf{64}:1104, 1996.

\bibitem{Jefimenko-99a}
O.~D. Jefimenko.
\newblock The {Trouton-Noble} paradox.
\newblock {\em J. Phys. A: Math. Gen.}, \textbf{32}:3755, 1999.

\bibitem{Jackson-torque}
J.~D. Jackson.
\newblock Torque or no torque? {Simple} charged particle motion observed in
  different inertial frames.
\newblock {\em Am. J. Phys.}, \textbf{72}:1484, 2004.

\bibitem{GS}
O.~W. Greenberg,  S.~S. Schweber.
\newblock Clothed particle operators in simple models of quantum field theory.
\newblock {\em Nuovo Cim.}, \textbf{8}:378, 1958.

\bibitem{Shirokov4}
A.~V. Shebeko,  M.~I. Shirokov.
\newblock Unitary transformations in quantum field theory and bound states.
\newblock {\em Phys. Part. Nucl.}, \textbf{32}:15, 2001.
\newblock http://www.arxiv.org/abs/nucl-th/0102037v1.

\bibitem{Close-Osborn}
F.~E. Close,  H.~Osborn.
\newblock Relativistic center-of-mass motion and the electromagnetic
  interaction of systems of charged particles.
\newblock {\em Phys. Rev. D}, \textbf{2}:2127, 1970.

\bibitem{Krajcik-Foldy}
R.~A. Krajcik,  L.~L. Foldy.
\newblock Relativistic center-of-mass variables for composite systems with
  arbitrary internal interactions.
\newblock {\em Phys. Rev. D}, \textbf{10}:1777, 1974.

\bibitem{Caprez}
A.~Caprez, B.~Barwick, H.~Batelaan.
\newblock A macroscopic test of the {Aharonov-Bohm} effect, 2007.
\newblock http://www.arxiv.org/abs/0708.2428v1.

\bibitem{Jackson}
J.~D. Jackson.
\newblock {\em Classical electrodynamics}.
\newblock J. Wiley and Sons, 3rd edition, 1999.

\bibitem{Boyer-Ahar}
T.~H. Boyer.
\newblock The paradoxical forces for the classical electromagnetic lag
  associated with the {Aharonov-Bohm} phase shift, 2005.
\newblock http://www.arxiv.org/abs/physics/0506180v1.

\bibitem{Boyer-2007-08}
T.~H. Boyer.
\newblock Comment on experiments related to the {Aharonov-Bohm} phase shift,
  2007.
\newblock http://www.arxiv.org/abs/0708.3194v1.

\bibitem{Boyer-2007}
T.~H. Boyer.
\newblock Unresolved classical electromagnetic aspects of the {Aharonov-Bohm}
  phase shift, 2007.
\newblock http://www.arxiv.org/abs/0709.0661v1.

\bibitem{Matteucci}
G.~Matteucci, D.~Iencinella, C.~Beeli.
\newblock The {Aharonov-Bohm} phase shift and {Boyer's} critical
  considerations: {New} experimental result but still an open subject?
\newblock {\em Found. Phys.}, \textbf{33}:577, 2003.

\end{thebibliography}

\end{document}